\documentclass[
 reprint,
superscriptaddress,
 amsmath,amssymb,
 aps
]{revtex4-1}

\usepackage[utf8]{inputenc}
\usepackage[T1]{fontenc}
\usepackage{mathptmx}
\usepackage{subfig}
\usepackage{tabularx}
\usepackage[italian,english]{babel}
\usepackage[dvipsnames]{xcolor}
\usepackage[version=4]{mhchem}
\usepackage{physics}
\usepackage{siunitx}
\usepackage{comment}
\usepackage{graphicx}
\usepackage{dcolumn}
\usepackage{bm}
\usepackage{amsmath}
\usepackage{mathtools, nccmath}
\usepackage{amsfonts}
\usepackage{amssymb}
\usepackage{xcolor}
\usepackage{bbold}

\renewcommand{\vec}{\bm}

\begin{document}

\preprint{APS/123-QED}

\title{Realization of $0-\pi$ states in SFIS Josephson junctions. The role of spin-orbit interaction and lattice impurities.}

\author{M. Minutillo}
\thanks{These two authors equally contributed to this work.}
\affiliation{Dipartimento di Fisica E. Pancini$,$ Università degli Studi di Napoli Federico II$,$ Monte S. Angelo$,$ via Cinthia$,$ I-80126 Napoli$,$ Italy}
\affiliation{CNR-SPIN$,$ UOS Napoli$,$ Monte S. Angelo$,$ via Cinthia$,$ I-80126 Napoli$,$ Italy}
\email{martina.minutillo@unina.it} 
\author{R. Capecelatro}
\thanks{These two authors equally contributed to this work.}
\affiliation{Dipartimento di Fisica E. Pancini$,$ Università degli Studi di Napoli Federico II$,$ Monte S. Angelo$,$ via Cinthia$,$ I-80126 Napoli$,$ Italy}
\author{P. Lucignano}
\affiliation{Dipartimento di Fisica E. Pancini$,$ Università degli Studi di Napoli Federico II$,$ Monte S. Angelo$,$ via Cinthia$,$ I-80126 Napoli$,$ Italy}

\begin{abstract}

Josephson devices with ferromagnetic barriers have been widely studied. Much less is known when the  ferromagnetic layer is insulating. In this manuscript we investigate the transport properties of superconductor-ferromagnetic insulator-superconductor (SFIS) junctions with particular attention to the temperature behavior of the critical current, that may be used as a fingerprint of the junction. We investigate the specific role of impurities as well as of possible spin mixing mechanisms, due to the spin orbit coupling. Transition between the $0$ and the $\pi$ phases can be properly tuned, thus achieving stable $\pi$ junctions over the whole temperature range, that may be possibly employed in superconducting quantum circuits.

\end{abstract}

\maketitle

\section{Introduction}
The supercurrent flowing across a Josephson Junction (JJ) is usually described by the sinusoidal relation $I = I_{c} \sin \varphi$, where $I_{c}$ is the critical current and $\varphi$ the phase difference across the junction. The Josephson energy associated to the latter reads: $E_J = \phi_0 I_c/(2\pi)\left(1-\cos{\varphi}\right)$ \cite{Tinkham, Barone_1982}  ($\phi_0$ is the superconducting flux quantum) implying that the  ground state energy, as well as the corresponding phase difference $\varphi$, depend on the sign of the supercurrent. 
In conventional JJs the ground state occurs at $\varphi = 0$ ($I_{c}>0$), by contrast in $\pi$ junctions the minimum energy corresponds to $\varphi = \pi$ ($I_{c}<0$ ). 
Superconductor- ferromagnet - superconductor JJs (SFS JJs) are promising platforms to implement $\pi$ junctions \cite{Buzdin2003, Golubov2004, Buzdin2005, Bergeret2005}. Indeed, as a consequence of the phase change of the Cooper pair wave function which extends from S to F layer due to the proximity effect, they show an oscillating behavior of the critical current as a function of the length of the F region \cite{LarkinFFLO,fulde_ferrell}, thus providing the necessary sign change of $I_{c}$ to switch from $0$ to $\pi$.

The so-called Josephson $\pi$-junctions ($\pi$-JJs) are currently subject to intense research activity due to their applicability as architectural elements for the development and the improvement of nanostructures. Indeed, they are considered to be very promising ingredients to engineer superconducting circuits, nanoelectronics, spintronics and quantum computing devices \cite{Ioffe2002, Ortlepp, Baselmans}. 

Among these applications the possible integration of $\pi$ junctions in quantum circuits for superconducting qubits is quite promising, in view of the  increased  robustness against noise and electromagnetic interference induced by magnetic field sources and a more compact and simple design, opening the way to scalable devices \cite{Blatter2001, Yamashita2005, Ioffe1999, Majer2002, Balashov2007,yamashita2020}. 
 
The $0$-$\pi$ transitions, between the $0$ and the $\pi$ state of a JJ, in SFS JJs has been experimentally measured varying the thickness of the ferromagnet in different kinds of samples \cite{Kontos2002, Blum2002, Robinson2006}. Moreover, these interesting platforms supplied important experimental evidence of temperature induced $0$-$\pi$ transitions \cite{Ryazanov2001, Sellier2004, Frolov2004}, characterized by a peculiar cusp in the $I_{c}(T)$ behavior at the transition temperature, thus giving the first chance to realize both the $0$ and $\pi$ state in a single device.

Even though SFS JJs are suitable for achieving $0-\pi$ transitions and $\pi$ states, these systems are affected by dissipative effects due to their coupling to the environment \cite{Zaikin1988, Kato2007}, leading to short decoherence times in Josephson quantum devices. To overcome this issue, $\pi$-JJs with a nonmetallic barrier result to be well suited for these kinds of applications. As matter of fact, transport properties through ferromagnetic insulator based JJs (SFIS JJs) have been investigated both theoretically \cite{Kawabata2010, Kawabata2010_2} and experimentally \cite{Senapati2011}, displaying temperature induced $0-\pi$ transitions together with unconventional $I_{c}(T)$ behaviors \cite{PRLTafuri, ahmad2021coexistence}. 

Therefore, the possibility to arrange an equilibrium superconducting phase difference of $\pi$ across JJ has been theoretically predicted also in this case \cite{Tanaka1997, Kawabata2010}. Moreover, experimental evidence of macroscopic quantum tunneling (MQT) phenomena in SFIS JJs has been provided, thus giving promise for their application in quantum hybrid circuits \cite{Massarotti2015, Ahmad2020}.

In this intriguing scenario, one of the most challenging issues is to find an effective way of controlling the occurrence of $0-\pi$ transitions in SFIS JJs, through a direct action on their $I_{c}(T)$ behavior. It is well known that tuning the exchange field of the barrier is an available technique to manipulate the critical current \cite{ahmad2021coexistence}, thus driving the switching between $0$, $0-\pi$ and $\pi$ regimes. However, this approach does not provide an easy engineering of ferromagnetic JJs based devices. Indeed, the exchange field is an intrinsic property of the magnetic barrier and, moreover, the current experimental procedures required for its manipulation may induce decoherence and magnetic noise. For this reason, it would be extremely useful to find alternative and more accessible mechanisms that can be exploited to drive the $0-\pi$ transition in such devices. 

In this paper, we recognize spin-mixing effects and lattice impurities as good candidates to approach this kind of task. Spin-orbit coupling (SOC) \cite{Bercioux_2015, Krive, Reynoso2008, BergeretTokatly2015, Yokoyama2014, Egger2009, Alidoust2015} have already been studied as a source of intriguing anomalous Josephson effect \cite{Campagnano_2015, Minutillo_2018, Alidoust2021, Pal2019, Brunetti2013, Buzdin2008, Konschelle2015, Nesterov2016}. Moreover, it has been investigated in previous works \cite{Asano2019}  to induce the $\pi-0$ transition in SFS JJs. However a thorough study of the SOC effect on the $I_{c}(T)$ behavior in these $0-\pi$ junctions is lacking. On the other hand, while previous works investigated the $0-\pi$ transition induced in SIS JJs by magnetic impurities \cite{Bulaevskii1977, Pal2018, Wang2012, Vavra2006}, the chance of controlling the $0-\pi$ transition by means of lattice non-magnetic impurities is not yet well explored.

In this article, we study the Josephson effect in SFIS JJs using a 2D Bogolioubov de Gennes (BdG) tight-binding model \cite{ ahmad2021coexistence, Furusaki1994, Asano2019, Asano2001}. The proximity effect in SFS junctions has widely been investigated with quasiclassical approaches (i.e. Eilenberger equations, \cite{Belzig1999, Lambert1998, Buzdin2005, Golubov2004, Bergeret2005}), however in our case we consider short tunnel junctions using the Gor’kov equation.

We investigate the role of spin-orbit coupling (SOC), on-site exchange field inhomogeneities and lattice impurities, in determining the transport properties of SFIS JJs. In particular, we extend the results in Ref.\cite{Asano2019} to the temperature dependence of $I_{c}$, by identifying the SOC as a tool to drive the switching between the $0-\pi$ and $0$ regimes. We  focus on $I_{c}$ because it is easily experimentally accessible and its behavior can encode interesting and direct information on the junction state.

Moreover, we analyze the peculiar role that impurities play in this respect, finding out the opportunity to reach a temperature stable $\pi$ regime, starting from a $0-\pi$ JJ, by increasing the strength of the impurity potential. Therefore, by controlling the latter, a well-defined phase difference of $\pi $ can be established between the two superconductors separated by a ferromagnetic insulator, which shows great stability over the whole temperature range.

In addition we present how the interplay between SOC and non-magnetic disorder may be exploited for the engineering of fully tunable $0-\pi$ JJs which can be switched between the $0$, $0-\pi$ and $\pi$ regimes.

Furthermore, since S/F hybrid systems are presumed to host triplet superconductivity induced by the proximity effect \cite{Golubov2004, Buzdin2005, Bergeret2005, Eschrig2007, Eschrig2008, Lofwander2010}, as supplementary analysis we also calculate the correlation functions in these SFIS JJs in the presence of SOC and impurities, both for s-wave and p-wave symmetry. Indeed, besides the short-range singlet and triplet pairings with total spin projection $S_{z}=0$, in our system we also find the equal spin-triplet long-range correlation pairs with total spin projection $S_{z}=\pm 1$ on the direction of the exchange field, due to the presence of SOC which allows the spin symmetry breaking at S/FI interfaces. In particular, we show that intensifying the impurities strength results in an enhancement of the odd-frequency correlations (i.e. s-wave equal-spin triplet and p-wave singlet, respectively).

Therefore, we identify SFIS JJs as sources of unconventional odd-frequency superconductivity and equal-spin triplet pairings when spin-mixing (SOC) and disorder effects are involved \cite{Bergeret2005, Eschrig2007, Eschrig2008, Lofwander2010, Linder_2019, Asano2019, Black_Schaffer_2012, Black_Schaffer_2020, Lothman_2020}, confirming the results in Refs.\cite{PRLTafuri, ahmad2021coexistence}.

The manuscript is structured as follows. In Sec.~\ref{model} we introduce our BdG model  hamiltonian \cite{ahmad2021coexistence, Furusaki1994, Asano2019}  and describe the recursive Green's function approach \cite{Asano2001, Furusaki1994}  used to calculate currents and correlation functions. In Sec.~\ref{results} and Sec.~\ref{correlazioni} we show a detailed investigation on the critical current as well as of the induced pairing functions. Sec.~\ref{conclusioni} summarizes our main findings. 
\section{Model}\label{model}
\subsection{Hamiltonian of the SFIS JJ}
\begin{figure}[t]
	\includegraphics[scale=0.235]{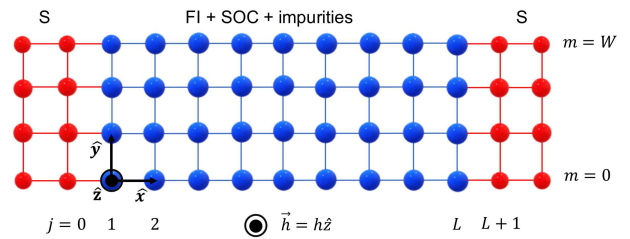}
	\caption[justified]{Schematic representation of the SFIS junction geometry with a ferromagnetic insulator barrier in the presence of spin-orbit coupling (SOC) and impurities. The exchange field $\vec{h}$ is taken parallel to the $z$-axis, thus perpendicular to the junction plane (i.e. the $xy$-plane).}   
	\label{Fig_1_model_JJ}
\end{figure}
In this section we illustrate the numerical calculation method used to study the transport properties in insulating ferromagnetic Josephson junctions (SFIS JJs), based on the recursive Green's function technique (RGF) \cite{Furusaki1994, Asano2019, Asano2001}.

We describe the SFIS JJ using a two dimensional lattice model, shown in Fig.\ref{Fig_1_model_JJ}, where $L$ is the length of the ferromagnetic insulator barrier, $W$ is the width of the junction, expressed in units of lattice sites, and $\vec{x}$ ($\vec{y}$) is the unit vector along the longitudinal (transverse) direction. Here, the vector $\vec{r}=j\vec{x}+m\vec{y}$ points a lattice site with $j=0,1,\ldots,L,L+1$ and $m=1,\ldots,W$. The Hamiltonian of the junction in the Nambu $\otimes$ spin space is given by
\begin{center}
	\begin{equation}
	\label{H_junction}
	\check{\mathcal{H}} = \sum_{\vec{r},\vec{r}'} \Psi^{\dagger}(\vec{r}) \begin{bmatrix} \hat{H}(\vec{r},\vec{r}') & \hat{\Delta}(\vec{r},\vec{r}') \\
	-\hat{\Delta}^{*}(\vec{r},\vec{r}') & - \hat{H}^{*}(\vec{r},\vec{r}')
	\end{bmatrix} \Psi(\vec{r}')
	\end{equation}
\end{center}
with
\begin{center}
	\begin{equation}
	\Psi(\vec{r}) = \left[ \psi_{\uparrow}(\vec{r}), \psi_{\downarrow}(\vec{r}), \psi^{\dagger}_{\uparrow}(\vec{r}), \psi^{\dagger}_{\downarrow}(\vec{r})  \right]^{T}.
	\end{equation}
\end{center}
$\psi^{\dagger}_{\alpha}(\vec{r})$ and $\psi_{\alpha}(\vec{r})$ are the field operators creating/annihilating an electron with spin $\alpha$ at lattice point $\vec{r}$.
Here, the symbols $\hat{.}$ and $\check{.}$ describe the $2 \times 2$ and $4 \times 4$ matrices in spin and Nambu$\otimes$spin spaces respectively.

In Eq.(\ref{H_junction}), $\hat{H}$ is the normal-state hamiltonian of the junction while $\hat{\Delta}$ describes the superconducting pairing potential. The former can be written as $\hat{H} = \hat{H}_{s} + \hat{H}_{FI} $, with $\hat{H}_{s}$ and $\hat{H}_{FI}$ referring to the S leads and FI barrier, respectively.

In Fig.\ref{Fig_1_model_JJ}, the S regions extend for $j < 1$ and $j > L$. $\hat{H}_{s}$ consists in a kinetic term (i.e. $\hat{H}_{s}=\hat{H}_{s}^{K}$) that reads:

\begin{center}
	\begin{align}
	\label{H_k_S}
	\hat{H}_{s}^{K}(\vec{r},\vec{r}')= & \lbrace -t_{s} \left( \delta_{\vec{r},\vec{r}'+\vec{x}} + \delta_{\vec{r}+\vec{x},\vec{r}'} \right) \hat{\sigma}_{0}\\ \notag
	&-t_{s} \left( \delta_{\vec{r},\vec{r}'+\vec{y}} + \delta_{\vec{r}+\vec{y},\vec{r}'} \right) \hat{\sigma}_{0}\\ \notag
	&- \left( 4 t_{s} - \mu_{s} \right) \delta_{\vec{r},\vec{r}'} \hat{\sigma}_{0} \rbrace \\ \notag
	& \times \left[\Theta \left( -j + 1 \right)+\Theta \left( j - L \right)\right] \\ \notag
	\end{align}
\end{center} 

where $t_{s}$ and $\mu_{s}$ are the hopping parameter and the chemical potential, respectively, and $\Theta$ is the Heaviside step-function. 
Here and in the followings, we indicate with $\hat{\sigma}_{0}$ and $\hat{\sigma}_{\nu}$ ($\nu = 1, 2, 3$) the unit and the Pauli matrices in the spin space, respectively.

In this work, we take the pairing potential $\hat{\Delta}$ different from zero only in the S leads, which, thus, vanishes inside the FI barrier. Here, $\hat{\Delta}$ is of spin-singlet s-wave symmetry and is expressed as

\begin{center}
	\begin{align}
	\label{Delta}
	\hat{\Delta}(\vec{r},\vec{r}') & =  \Delta \delta_{\vec{r},\vec{r}'} \, i \, \hat{\sigma}_{2}\\\notag
	&\times \left[ \Theta \left( -j + 1 \right) e^{i\phi_{L}} + \Theta \left( j - L \right) e^{i\phi_{R}} \right],
	\end{align}
\end{center}
where $\phi=\phi_{L}-\phi_{R}$ defines the phase difference across the junction and $\phi_{L}$ $(\phi_{R})$ is the phase in the left (right)-hand side superconductor. In our model, the order parameter $\Delta$ is constant in the leads and it is not derived from self-consistent calculations. Further, we assume that there is no disorder in the superconductors.

The FI barrier extends from $j=1$ to $j=L$ and its hamiltonian consists of four terms

\begin{center}
	\begin{equation}
	\label{H_FI}
	\hat{H}_{FI}= \hat{H}_{FI}^{K} + \hat{H}_{FI}^{SOC} + \hat{H}_{FI}^{ex}+\hat{H}_{FI}^{i} .
	\end{equation}
\end{center}

$\hat{H}_{FI}^{K}$ is the kinetic energy,
\begin{center}
	\begin{align}
	\label{H_k_FI}
	\hat{H}_{FI}^{K}(\vec{r},\vec{r}')= & \lbrace -t_{FI} \left( \delta_{\vec{r},\vec{r}'+\vec{x}} + \delta_{\vec{r}+\vec{x},\vec{r}'} \right) \hat{\sigma}_{0}\\ \notag
	&-t_{FI} \left( \delta_{\vec{r},\vec{r}'+\vec{y}} + \delta_{\vec{r}+\vec{y},\vec{r}'} \right) \hat{\sigma}_{0}\\ \notag
	&- \left( 4 t_{FI} - \mu_{FI} \right) \delta_{\vec{r},\vec{r}'} \hat{\sigma}_{0}\rbrace \\ \notag
	& \times \Theta \left( j \right) \Theta \left( L + 1 - j \right). \\ \notag 
	\end{align}
\end{center} 

The spin-orbit coupling (SOC) is

\begin{center}
	\begin{align}
	\label{H_SOC}
	&\hat{H}_{FI}^{SOC}(\vec{r},\vec{r}')= i \alpha \left[ \left\lbrace \delta_{\vec{r},\vec{r}'+\vec{x}} - \delta_{\vec{r}+\vec{x},\vec{r}'} \right\rbrace \hat{\sigma}_{2} \right. \\\notag
	&- \left.  \left\lbrace \delta_{\vec{r},\vec{r}'+\vec{y}} - \delta_{\vec{r}+\vec{y},\vec{r}'} \right\rbrace \hat{\sigma}_{1} \right] \Theta \left( j \right) \Theta \left( L + 1 - j \right) . \\ \notag
	\end{align}
\end{center} 

The exchange potential is

\begin{center}
	\begin{equation}
	\label{H_{ex}}
	\hat{H}_{FI}^{ex}(\vec{r},\vec{r}') = - \vec{h}' \cdot \boldsymbol{\sigma} \delta_{\vec{r},\vec{r}'} \Theta \left( j \right) \Theta \left( L + 1 - j \right) ,
	\end{equation}
\end{center} 

where $\boldsymbol{\sigma}$ is the vector of the Pauli matrices $\left( \hat{\sigma}_{1}, \hat{\sigma}_{2}, \hat{\sigma}_{3} \right)$. \\
The on-site random impurity potential is

\begin{center}
	\begin{equation}
	\label{H_i}
	\hat{H}_{FI}^{i}(\vec{r},\vec{r}') = v_{\vec{r}} \, \hat{\sigma}_{0} \, \delta_{\vec{r},\vec{r}'} \Theta \left( j \right) \Theta \left( L + 1 - j \right) \;  \; .
	\end{equation} 
\end{center}

In the above equations, $t_{FI}$  is the hopping integral among nearest-neighbor lattice sites (in the followings we omit the subscript FI for simplicity, i.e. we take $t_{FI} = t$), $\mu_{FI}$ the Fermi energy, $\alpha$ the amplitude of the spin-orbit interaction, $v_{\vec{r}}$ the on-site random impurity potential strength, uniformly distributed in the range $-V_{imp} \leq v_{\vec{r}} \leq V_{imp}$. Finally, the exchange field is assumed to be slightly disordered and is modeled as $\vec{h}' = \vec{h} + \vec{\delta_{h}}$, where $\vec{\delta_{h}}$ are small on-site fluctuations given randomly in the range $-h/10 \leq \delta_{h} \leq h/10$ (along the $\vec{h}$-direction).
In this work, since the junction plane coincides with the $xy$-plane, the exchange field $\vec{h}'$ is always taken in the perpendicular direction, $\vec{h}'=h'\vec{z}$ (along the $z$-direction).

\subsection{Transport properties of the SFIS JJ}
In this context, the transport properties of the JJ can be derived from the  Green’s function (GF) $\check{G}$ of the barrier. Its Matsubara representation is given by $\check{G}_{\omega_{n}}(\vec{r},\vec{r}')$, where  $\vec{r}=j\vec{x}+m\vec{y}$ and $\vec{r}'=j'\vec{x}+m'\vec{y}$ run over all the possible lattice site indices $j,j'=0,1,\ldots,L,L+1$ and $m,m'=1,\ldots,W$. The barrier Matsubara GF is, thus, a $4WL\times4WL$ matrix in the Nambu$\otimes$spin space, whose blocks with fixed ($\vec{r}$, $\vec{r}'$) can be numerically calculated by solving the Gor'kov equation
\begin{center}
	\begin{align}
	\label{Gorkov_eq}
	&\left[ i \omega_{n} \hat{\tau}_{0} \hat{\sigma}_{0} - \sum_{\vec{r}_{1}} 
	\begin{pmatrix} 
	\hat{H}(\vec{r},\vec{r}_{1}) & \hat{\Delta}(\vec{r},\vec{r}_{1}) \\
	-\hat{\Delta}^{*}(\vec{r},\vec{r}_{1}) & -\hat{H}^{*}(\vec{r},\vec{r}_{1})
	\end{pmatrix}   \right]  \\\notag
	&\times \check{G}_{\omega_{n}}(\vec{r}_{1}, \vec{r}')  = \hat{\tau}_{0} \hat{\sigma}_{0} \delta(\vec{r}-\vec{r}'),
	\end{align}
\end{center}
where $\omega_{n}=(2n+1)\pi T$ is the fermionic Matsubara frequency and $T$ is the temperature.  Here and in the followings, $\hat{\tau}_{0}$ and $\hat{\tau}_{\nu}$ ($\nu=1,2,3$) are the analogous of the unit and Pauli's matrices in the Nambu space, respectively. 

We solve the Gor’kov Eq.(\ref{Gorkov_eq}) by applying the recursive Green's function technique (RGF) \cite{Lee_1981, Ando_1991, Datta_1995, Furusaki1994, Asano2019, Asano2001} that consists in dividing the 2D lattice along the $\vec{x}$-direction in transverse stripes, and recursively calculating the GFs at each stripe of the barrier, starting from the two superconducting leads (see Appendix B for more details). 

For this reason, we introduce the GF in Nambu$\otimes$spin space $\check{G}_{j, j}$ of the stripe $j$ along the $x$-direction inside the barrier. It is a $4W\times4W$ matrix, where $W$ is the number of lattice sites in each stripe. It can be visualized as a $2\times 2$ block matrix in the Nambu space, where each block consists in a $2W\times2W$ sub-matrix:  
\begin{center}
	\begin{equation}
	\label{Nambu_spin_GF_4W}
	\check{G}_{j, j} = \check{G}_{\omega_{n}}(\vec{r},\vec{r}') = \begin{bmatrix}
	\hat{G}_{\omega_{n}} (\vec{r},\vec{r}') & \hat{F}_{\omega_{n}} (\vec{r},\vec{r}') \\
	-\hat{F}^{*}_{\omega_{n}} (\vec{r},\vec{r}') & -\hat{G}^{*}_{\omega_{n}} (\vec{r},\vec{r}')
	\end{bmatrix} ,
	\end{equation}
\end{center}
where $\vec{r}=j\vec{x}+m\vec{y}$, $\vec{r}'=j\vec{x}+m'\vec{y}$ with $j$ fixed. 
The off-diagonal terms of the matrix in the right-hand side of Eq.(\ref{Nambu_spin_GF_4W}) are the so-called \emph{anomalous Green’s functions} $\hat{F}_{\omega_{n}}$, that describe the superconducting pair correlations. Notice that, here and in the following expressions, the 'checks' $\check{.}$ indicate the full $4W \times 4W$ matrices in Nambu$\otimes$spin space, whereas $\hat{.}$ is used for $2W \times 2W$ matrices in the spin space. 

In order to compute the GF of the whole barrier, we start from the surface GFs of the two superconducting leads S, as they are reported in Ref.\cite{Furusaki1994}, and recursively calculate the GF $\check{G}_{j, j}$ at each stripe $j$ inside the FI barrier by connecting it to the leads with the RGF technique (see Appendix B for more details). 

The barrier stripes can be connected (to the leads and to each other) by using the hopping matrices $\check{T}^{\pm}$, given by 
\begin{center}
	\begin{equation}
	\label{T_matrix}
	\check{T}^{\pm}= \begin{pmatrix}
	-t & \mp\alpha & 0 & 0 & \ldots \\
	\pm\alpha & -t & 0 & 0 & \ldots \\
	\ldots & \ldots & \ldots & \ldots & \ldots \\
	0 & 0 & \ldots & t & \pm\alpha \\
	0 & 0 & \ldots & \mp\alpha & t
	\end{pmatrix} \, ,
	\end{equation}
\end{center}
involving the hopping and spin-orbit coupling along the $x$-direction.
Since we consider nearest-neighbors hopping, only adjacent stripes can be connected by the $\check{T}^{\pm}$ matrices (i.e. the $j$-th stripe is linked to the $j-1$-th and $j+1$-th).

In particular, with the RGF method we derive the GFs connecting two adjacent stripes (namely the $j$-th and $j+1$-th), $\check{G}_{j,j+1} =\check{G}_{\omega_{n}} (\vec{r},\vec{r}' + \vec{x})$ and $\check{G}_{j+1,j} =\check{G}_{\omega_{n}} (\vec{r} + \vec{x}, \vec{r}')$ with $\vec{r}=j\vec{x}+m\vec{y}$, $\vec{r}'=j\vec{x}+m'\vec{y}$. From these latter, the Josephson current, at finite temperature and given position $j$ in FI, can be computed as follows

\begin{center}
	\begin{equation}
	\label{Josephson_curr}
	I(j) = - \dfrac{i e}{2} T \sum_{\omega_{n}}  Tr \left[ \hat{\tau}_{3} \check{T}_{+} \check{G}_{j,j+1}- \hat{\tau}_{3} \check{T}_{-} \check{G}_{j+1,j}   \right] ,
	\end{equation}
\end{center}
 where \textit{Tr} stands for the trace over the Nambu$\otimes$spin space. In the above equation, the summation over the Matsubara frequencies is performed.
 
 Furthermore, by taking the off-diagonal elements of the stripes GFs ($\check{G}_{j,j}$) in the right-hand side of Eq.(\ref{Nambu_spin_GF_4W}), $\hat{F}_{\omega_{n}}(\vec{r},\vec{r}')$ with $\vec{r}=\vec{r}'=j\vec{x}+m\vec{y}$ (with $j$ fixed), we can derive the four pairing components with s-wave symmetry at each stripe $j$ along the $\vec{x}$-direction:
 \begin{center}
 	\begin{equation}
 	\label{s_wave_correlations}
 	\dfrac{1}{W} \sum_{\omega_{n}}\sum_{m=1}^{W} \hat{F}_{\omega_{n}}(\vec{r},\vec{r}) = \sum_{\nu=0}^{3} f_{\nu}(j) \hat{\sigma}_{\nu} \; i  \; \hat{\sigma}_{2} \; ,
 	\end{equation}
 \end{center} 
 where $f_{0}$ is the spin-singlet component and $f_{\nu}$ with $\nu=1,2,3$ are the spin-triplet components.
 
 Analogous considerations can be applied to the GFs connecting the $j$-th stripe with its neighbors $j \pm 1$ (i.e. $\check{G}_{j,j+1}$, $\check{G}_{j+1,j}$, $\check{G}_{j,j-1}$, $\check{G}_{j-1,j}$), from which we can calculate the odd-parity pairing functions:
 
 \begin{gather}
 \notag \dfrac{1}{4W} \sum_{\omega_{n}}\sum_{m=1}^{W} \hat{F}_{\omega_{n}}\left( \vec{r}+\vec{x},\vec{r}\right)  + \hat{F}_{\omega_{n}}\left( \vec{r},\vec{r}-\vec{x}\right)  \\  - \hat{F}_{\omega_{n}}\left(\vec{r},\vec{r}+\vec{x}\right)  
 -\hat{F}_{\omega_{n}}\left( \vec{r}-\vec{x},\vec{r}\right)=  \label{p_wave_correlations} \\
 \notag\sum_{\nu=0}^{3} f_{\nu}\left( j\right)  \hat{\sigma}_{\nu} \; i  \; \hat{\sigma}_{2} \, , 
 \end{gather}
 that give rise to p-wave superconductivity.
 
 Making explicit the term in right hand side of Eqs.(\ref{s_wave_correlations}) (Eq.(\ref{p_wave_correlations})), we can rewrite the s-wave (p-wave) pairing components as
 \begin{center}
 	\begin{equation}
 	\label{corrs_singlet_triplet}
 	\begin{cases}
 	f_{0} = \dfrac{f_{\uparrow \, \downarrow} - f_{\downarrow \, \uparrow}}{2} \\
 	\\
 	f_{3} = \dfrac{f_{\uparrow \, \downarrow} + f_{\downarrow \, \uparrow}}{2}  \\
 	\\
 	f_{1} = \dfrac{f_{\downarrow \, \downarrow} - f_{\uparrow \, \uparrow}}{2} \\
 	\\
 	f_{2} = \dfrac{f_{\uparrow \, \uparrow} + f_{\downarrow \, \downarrow}}{2 \, i}  ,
 	\end{cases}
 	\end{equation}
 \end{center}
 from which we extract the standard spin correlation functions, $f_{0}$, $f_{3}$, $f_{\uparrow}$ (that is $f_{\uparrow\uparrow}$) and $f_{\downarrow}$ (that is $f_{\downarrow\downarrow}$).

In the following we report the choice of the model parameters used for these calculations. Henceforth, we adopt units with $\hbar=c=k_B=1$, where $c$ is the speed of light and $k_B$ is the Boltzmann constant. 

All the energies are, thus, scaled by $t$ and the magnitude of the spin-orbit coupling $\alpha$ is scaled by $ta$, where the lattice constant is set $a=1$, while the Josephson current is calculated in units of $I_{0}=e\Delta$.
Further, we fix several parameters as $t_{s} = t = 1$, $\mu_{FI} = 0$, $\mu_{s} = 3$, $\Delta = 0.005$. The chosen chemical potential mismatch at FI and S interfaces allows to describe the insulating regime in our model.
 
In the numerical simulations, both the temperature dependence of the critical Josephson current (i.e. $I_{c}(T)$) and correlations functions are averaged over $N_{s}$ samples with different random impurity configurations. We choose $N_{s}=80-100$ for the $I_{c}(T)$ curves and $N_{s}=300$ for the pairing correlation functions. In particular, the ensemble average of the Josephson current (Eq.\ref{Josephson_curr}) over a number of different samples is obtained as: $< I > = \frac{1}{N_{s}} \sum_{n = 1}^{N_{s}} I_{n}$, where $I_{n}$ is the Josephson current in the $n$-th sample. Then, evaluating the average Josephson current by varying $\phi$ in the range from $0$ to $\pi$ we obtain the average current-phase relation (CPR) at fixed $T$ (i.e. $I(\phi,T)$). Finally, the averaged critical current $I_{c}(T)$ is estimated from the CPR at different temperatures between $0$ and $T_{c}$, by taking its maximum in absolute value ($I_{c}(T)=\max_{\phi}[|I(\phi,T)|]$). In this work, each $I_{c}(T)$ curve together with the corresponding CPRs ($I(\phi,T)$) has been normalized to the maximum value of the critical current w.r.t. the temperature, i.e. $I_{max}=\max_{T}[I_{c}(T)]$.  

\section{Josephson critical current $I_{c}(T)$} \label{results}
\subsection{Clean limit: the role of spin-orbit coupling (SOC).  Switching between $0$  and $\pi$ state.}

\begin{figure}[h!]
	\includegraphics[scale=0.35]{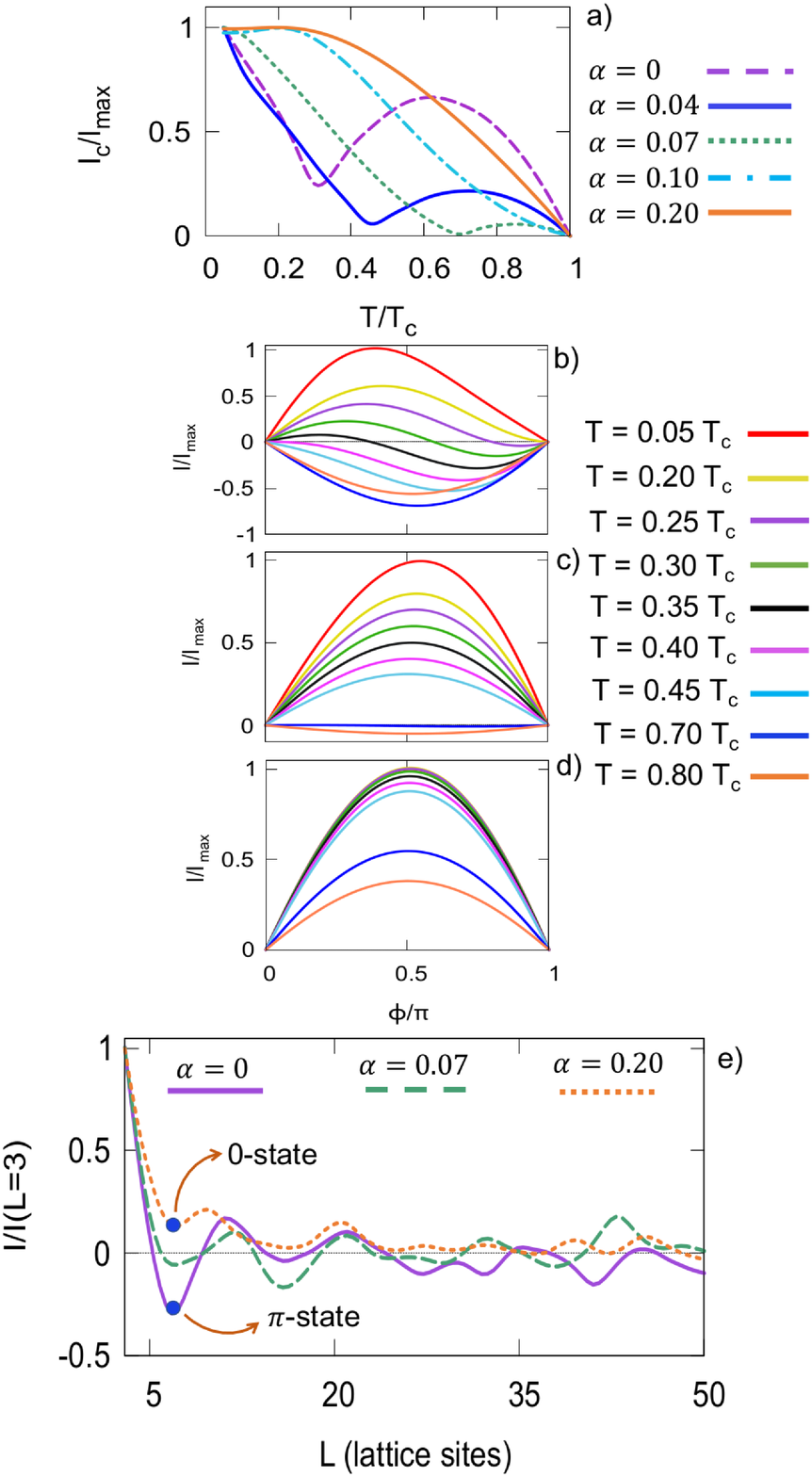}
	\caption[]{Effect of SOC increasing on the $I_{c}(T)$ (a) and current-length relation (CLR) (e). We set the exchange field $h=0.45$ in (a-e). In (a-d), the dimensions of the lattice are $L=8$ and $W=32$, along the $x$ and $y$-direction, respectively. In (b-d), the CPRs relative to the simulations in (a) for $\alpha=0,0.07,0.2$ are shown.  In (e), the CLR is computed by varying the length of the junction from $L=3$ to $L=50$, with $\phi=\pi/2$, $T=0.1T_{c}$ and $W=32$, for $\alpha=0,0.07,0.2$. The CLR curves in (e) are normalized to the value of their current at $L=3$.}   
	\label{Fig_2_alpha_var}
\end{figure}

In order to investigate  efficient tools for controlling the transition between $0$, $0-\pi$ and $\pi$ regimes, we focus on how spin mixing and disorder effects can be exploited for this task. In this section, we consider the case of a clean ($V_{imp}=0$) $0-\pi$ SFIS JJ which presents temperature induced $0-\pi$ transitions and analyze the effect of SOC on its $I_{c}(T)$ behavior. 

Here and in the following, we consider the short junction regime, typical of tunnel junctions, where the length of the FI barrier is fixed at $L = 8$ sites. The other model parameters are: $W = 32$, $h=0.45$, $V_{imp}=0$, $\delta_{h}=0$. In Fig.\ref{Fig_2_alpha_var} (a) we can see that, starting from a $0-\pi$ transition corresponding to $\alpha = 0$, we recover the Ambegaokar-Baratoff (AB) behavior \cite{AB} for $\alpha = 0.20$. For larger $\alpha$ the critical current is always of AB type. We can also notice that the second lobe of the $I_{c}(T)$ curve is reduced in height and its (cusp-like) minimum is shifted toward higher temperatures when $\alpha$ increases, until the $I_{c}(T)$ behavior visibly changes and it completely disappears in the AB regime. In Figs.\ref{Fig_2_alpha_var} (b-d), the characteristic current-phase relation (CPR) corresponding to $I_{c}(T)$ curves for $\alpha = 0, \, 0.07, \, 0.20$ are shown. We can notice that the $0-\pi$ transition is well evident for $\alpha = 0$ and only slightly appreciable for $\alpha = 0.07$, while it is completely washed out for $\alpha = 0.20$. Hence, by increasing $\alpha$, we induce a shift in the transition temperature $T_{0-\pi}$ towards higher values, until the $0-\pi$ transition cancels out. Further, we observe visible contributions due to the second and higher order harmonics in the CPRs at $T_{0-\pi}$, when the first order harmonic appears strongly weakened \cite{Goldobin_2007, Konschelle2008}. Our results confirm the fact that SOC stabilizes the $0$ state rather than the $\pi$ state \cite{Asano2019} in these devices.

To better illustrate this mechanism, in Fig.\ref{Fig_2_alpha_var} (e) we show the critical Josephson current as a function of the length of the ferromagnetic-insulator layer (with $L$ from $L =3$ to $L =50$), i.e. current-length relation (CLR) . In these simulations we used $T = 0.1 T_{c}$, $\phi = \pi/2 $ and $W = 32$ sites and $y$ direction, respectively. Furthermore, we set $h = 0.45$, $\alpha = 0, 0.07, 0.20 $ and a uniform distribution of the exchange field in the 2D lattice is assumed ($\delta_{h}=0$). 

The change in sign of Josephson current, indicating the corresponding $0-\pi$ transitions at fixed lengths of the FI layer due to the presence of the exchange potential, is more frequent in the cases with $\alpha=0$ and $\alpha=0.07$. The effect of SOC increasing is the suppression of the above-mentioned $0-\pi$ transitions and consequently a mostly always positive Josephson current for $\alpha=0.20$. Indeed, the SOC produces a shift of the CLR curve from lower to higher current values; negative critical currents are representative of $\pi$ states while, when $I_{c}$ becomes positive (at fixed length L), the $\pi-0$ transition occurred and the system reaches the $0$ final-state under the SOC growth. Taking $h$ and L fixed, there is no possibility that, by further increasing $\alpha$, the system will return to the $\pi$ state experiencing another $0-\pi$ transition. 

A qualitative explanation of this effect follows. The short-range spin-triplet component $S_{z}= 0$ appears in S/FI systems due to exchange field breaking time-reversal symmetry in F. In the presence of spin-orbit interaction, the spin-mixing effect at S/FI interfaces allows for long-range equal spin-triplet components $S_{z}=\pm 1$ inside the ferromagnetic region. The $S_{z}=0$ components (singlet and zero-spin triplet) show an oscillating behavior due to the different phase shifts acquired by the up and down-spin electrons of the Cooper pair, as they propagate in F, while the $S_{z}=\pm 1$ pairing components show a long-range decay, since the exchange field $h$ has the same effect on the two equal spin electrons \cite{LarkinFFLO, fulde_ferrell, Bergeret2005, Bergeret2012}. In the long junction limit all the pairing functions decay exponentially over the thermal coherence length $\xi_{T} = \frac{v_{F}}{2 \pi T}$, whereas the oscillation period of the zero-spin pairing components is given by $\xi_{h} = \frac{v_{F}}{2 h}$ \cite{Eschrig2007, Eschrig2008, Buzdin2005, Asano2019}. Therefore, for $x >> \xi_{T}$, heuristically, the Josephson current can be considered as consisting of two contributions:

\begin{gather}
\label{total_current}
I  \sim \,  I_{S_{z=\pm 1}} \, e^{-x/\xi_{T}} + I_{S_{z=0}} \, e^{-x/\xi_{T}} \,  \cos\left(\frac{x}{\xi_{h}}\right) \, , \notag
\end{gather} 
where $I_{S_{z=0}}$ and $I_{S_{z=\pm 1}}$ are the amplitudes of the opposite spin and parallel spin components, respectively.Increasing the SOC results in an enhancement of the non-oscillating part of the Josephson current, whose superposition with the oscillating term produces, in turn, an enlarged total current and, thus, may prevent that $I_{c}$ vanishes in the $0-\pi$ transition.

Ultimately, our results lead to regarding SOC as a useful tool for driving the evolution of the $I_{c}(T)$ of SFIS JJs from $0-\pi$ to $0$ regime. 

\subsection{The dirty regime: the role of impurities in the formation of $\pi$-JJs }
\begin{figure}[h!]
	\includegraphics[scale=0.35]{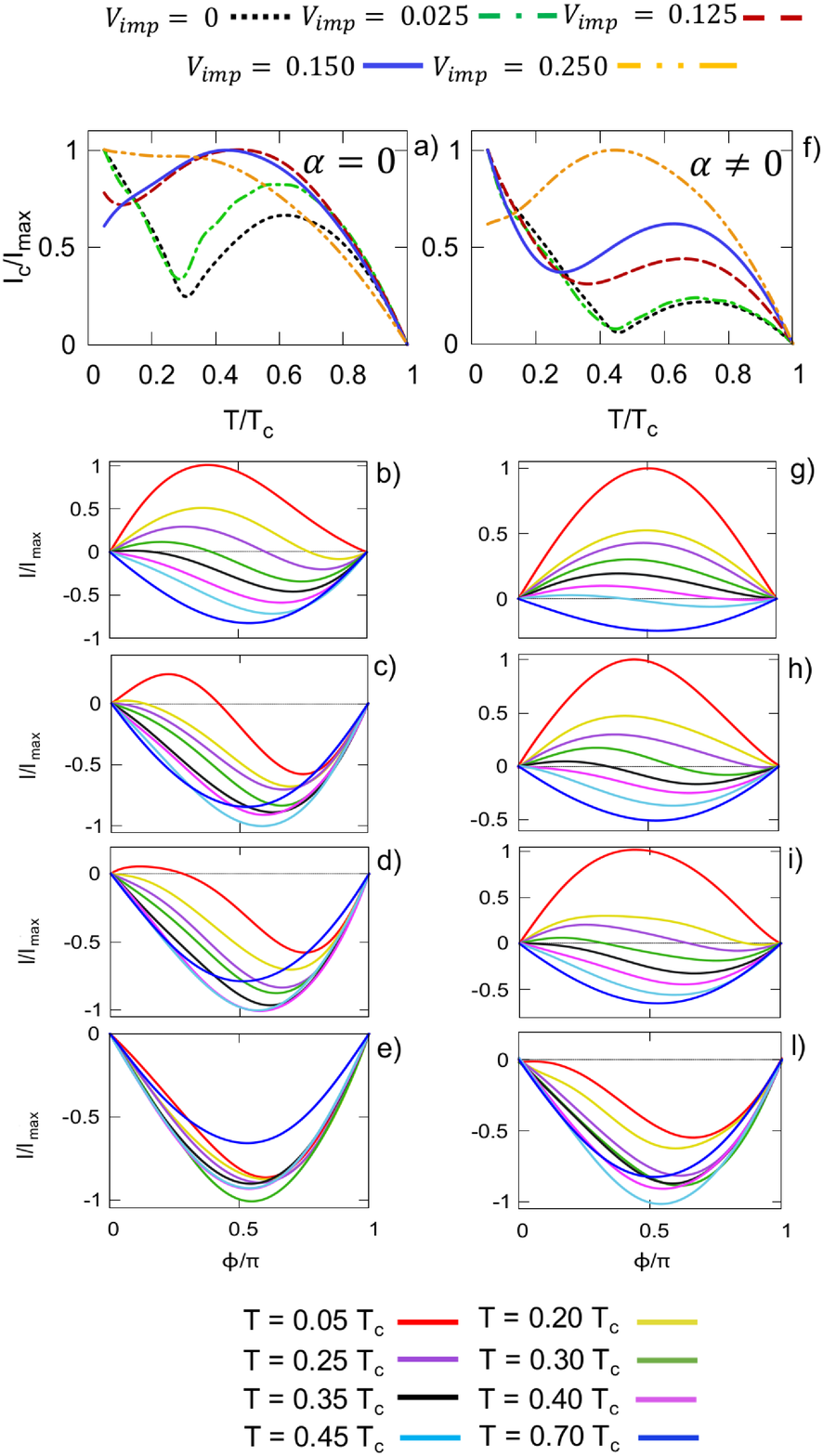}
	\caption[]{Effect of increasing $V_{imp}$ in the cases without and with SOC (a, f). Corresponding calculated CPRs (b-e, g-l) and formation of $\pi$ state (see text for details on the used parameters).} 
	\label{Fig_3_Ic_vimp}
\end{figure}

It is legitimate to ask under which conditions the Junction displays  a  stable $\pi$ state over the whole temperature range, similar to what happens in the case of the $0$ state with strong SOC. Real systems are affected by the unavoidable presence of impurities and in this section, we show that interesting features for the existence of $\pi$ states in SFIS JJs can be detected in the presence of non-magnetic impurities, modeled as scalar on-site potentials.

In what follows we focus on the effect of disorder on the $I_{c}(T)$ curves exhibiting a $0-\pi$ transition and on the corresponding CPR. In particular, firstly we consider the case of a SFIS JJ in the presence of disorder, secondly we analyze the combined effect of impurities and SOC on the $I_{c}(T)$ of these devices. As a matter of fact, the influence of increasing $V_{imp}$ results in different scenarios.

In the following simulations, we use $L=8$, $W = 32$ , $\alpha = 0.04 $ (for the case with SOC) and four different values of $V_{imp}= 0.025, 0.125, 0.150, 0.250$. Moreover, here, small on-site exchange field fluctuations are considered ($\delta_{h} \neq 0$) to model a more realistic scenario in which the exchange field may be non-uniform in the whole barrier.
In Fig.\ref{Fig_3_Ic_vimp} (a) the $I_{c}(T)$ curve at $\alpha = 0 $ is calculated as the impurity potential increases. We notice that for high values of $V_{imp}$ the system changes its $I_{c}(T)$ behavior leaving the $0-\pi$ regime and reaching a stable $\pi$ state almost over the whole temperature range, as it is evident in the corresponding CPRs (Figs.\ref{Fig_3_Ic_vimp} (c-e)). Precisely, for $V_{imp} = 0.025$ (green curve) both the maximum of the second lobe and the dip of the $0-\pi$ transition settle at higher values of current with respect to the clean regime (black line in Fig.\ref{Fig_3_Ic_vimp} (a)). Consequently, the clear effect of increasing $V_{imp}$ is the filling of the minimum of the $0-\pi$ transition and its shifting toward very low temperatures (red and blue curves), leading to the AB-like behavior for the highest disorder configuration (orange curve), corresponding to a pure $\pi$ regime (Fig.\ref{Fig_3_Ic_vimp} (e)). The presence of impurities seems to be a driving force for the conversion of a $0-\pi$ JJ into a pure $\pi$ one. As it is shown, in the absence of SOC the realization of an almost pure $\pi$-JJ is feasible even for small values of the impurity potential (Fig.\ref{Fig_3_Ic_vimp} (c, d)).  

In the presence of SOC (Fig.\ref{Fig_3_Ic_vimp} (f)) the clean $I_{c}(T)$ curve (black line) exhibits a $0-\pi$ transition occurring at $T \sim 0.45 T_{c}$, characterized by a lower value of the current in the $\pi$ state. In this case, we can notice that enhancing the impurity strength produces a more gradual filling of the $0-\pi$ dip together with its broadening (Fig.\ref{Fig_3_Ic_vimp} (f), red and blue curves), shifting it toward higher critical current values and lower temperatures. When passing from the $0-\pi$ regime to the $\pi$ one, with $\alpha\neq 0$, the system also shows a peculiar plateau region in the $I_{c}(T)$ extended over a wide range of temperatures, for intermediate values of $V_{imp}$ (Fig.\ref{Fig_3_Ic_vimp} (f), red curve).

The combined effect of SOC and impurities allows to stabilize the $0-\pi$ transition over a wide range of temperatures. However, for $V_{imp} = 0.250$, we observe a sharp change in the $I_c(T)$ behavior, where neither the plateau nor the $0-\pi$ dip are no longer visible, suggesting that the $0-\pi$ transition may occur at very low temperatures and that the pure $\pi$ regime may be reached at larger values of $V_{imp}$.

We may further analyze the system response to the presence of disorder by looking at the  CPRs (Figs.\ref{Fig_3_Ic_vimp} (b-e), (g-l)), corresponding to the $I_c(T)$ curves in Figs.\ref{Fig_3_Ic_vimp} (a) and (b), respectively.
For the first scenario (Fig.\ref{Fig_3_Ic_vimp} (a, b-e)), the increase of $V_{imp}$ produces strong modifications in the CPRs, characterized by an enhanced contribution of higher order harmonics at low temperatures, reflecting the lowering of the $0-\pi$ transition temperature ($T_{0-\pi}$). Further, for $V_{imp} = 0.250$ the CPRs are opposite in sign with respect to the typical $\sin \phi$ behavior, indicating that a phase difference of $\pi$ is established across the junction.

In Fig.\ref{Fig_3_Ic_vimp} (f, g-l), the CPRs in the case of $\alpha \neq 0$ are presented. Here, the $0-\pi$ transition gradually moves to lower temperatures as $V_{imp}$ increases, until the junction is totally $\pi$ for the highest value of $V_{imp}$ (Fig.\ref{Fig_3_Ic_vimp} (l)).

\begin{figure}[h!]
	\includegraphics[scale=0.30]{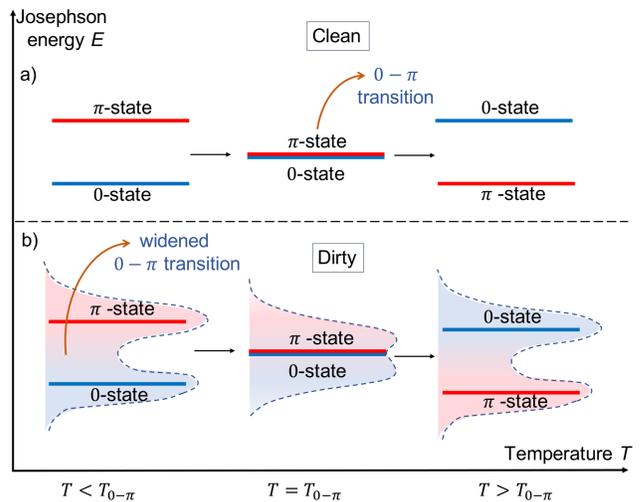}
	\caption[]{Schematic representation of $0$ and $\pi$ energy levels in the clean (a) and dirty (b) regime. In the latter case, the broadening in energy, due to the presence of on-site non-magnetic impurities, is shown.} 
	\label{Fig_4_broad_levels}
\end{figure}

In Fig.\ref{Fig_3_Ic_vimp} (f) we observe  that the influence of the SOC on the dirty SFIS JJs consists in stabilizing the $0-\pi$ regime even for moderately high values of the impurity potential. Here, the result of the coexistence of two competing effects, namely the spin-orbit and the disorder, is noticeable. Indeed, as for the clean regime, also in the dirty case the SOC tends to bring the system toward the $0$ state; whereas the non-magnetic on-site impurities encourage it to turn toward the $\pi$ state. This results in the slowdown of the switching from $0-\pi$ to $\pi$ $I_{c}(T)$ behavior, which, thus, takes place more gradually as $V_{imp}$ is enlarged. 
For this reason, the system goes through an intermediate regime involving a widened $0-\pi$ transition characterized by the plateau in the critical current, before reaching the $\pi$ regime. In this situation, we can better visualize how the impurities drive the transformation of the $I_{c}(T)$ from that of a $0-\pi$ JJ to the one of a $\pi$-JJ. Indeed, this mechanism is only slightly perceivable when $\alpha=0$ and the entire process happens almost suddenly.
We can provide a qualitative picture of this phenomenon in the following.
In the clean case, for temperatures lower than the $0-\pi$ transition one ($T < T_{0-\pi}$) the lower energy level is the $0$ state. When $T = T_{0-\pi}$ the $0$ and $\pi$ energy levels are coinciding; finally, for $T > T_{0-\pi}$ the $0-\pi$ transition has occurred and the Josephson energy minimum is reached at $\phi = \pi$ (Fig.\ref{Fig_4_broad_levels} (a)).
On the other hand, the presence of on-site non-magnetic impurities produces a broadening in energy (and, therefore, in temperature) of the $0$ and $\pi$ energy levels (Fig.\ref{Fig_4_broad_levels} (b)). The latter, becoming wider, give rise to an overlap region in which the JJ comes to be in a "hybrid $0-\pi$ state" over a more or less extended range of temperatures. When the impurities strength is enhanced, the overlapping between the levels grows together with the probability that, for $T > T_{0-\pi}$, the system prefers to stabilize in the lower $\pi$ energy state.

\begin{figure*}
 	\includegraphics[scale=0.50]{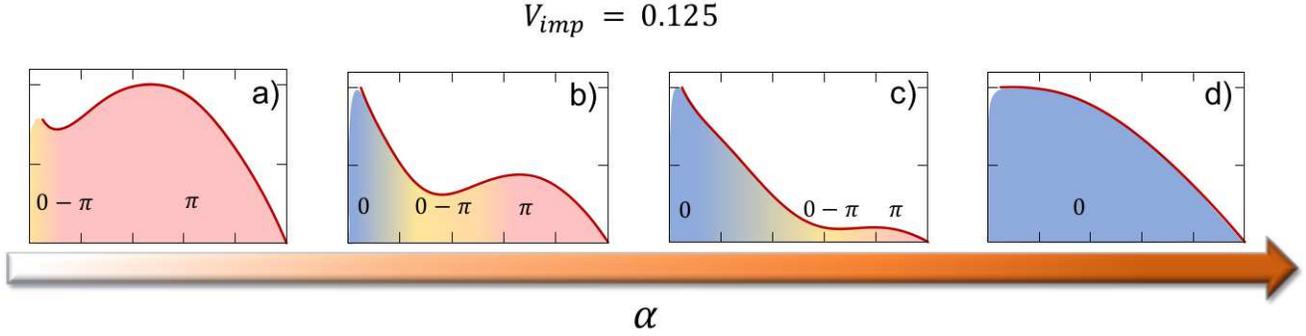}
 	\caption[]{Effect of SOC strength ($\alpha$) increasing on the $I_{c}(T)$ behavior, at fixed value of the impurity potential ($V_{imp}=0.125$). The $I_{c}(T)$ curves are calculated for the following values of $\alpha=0, \, 0.04, \, 0.07, \, 0.20$. The $I_{c}(T)$ curves in (a) and (b) correspond to the ones in Fig.\ref{Fig_3_Ic_vimp} (a) and (f), respectively, for $V_{imp}=0.125$.} 
 	\label{Fig_5_alpha_freccia}
 \end{figure*}

Finally, we illustrate, in Fig.\ref{Fig_5_alpha_freccia}, the possibility to build up a controllable device that can host all the $I_{c}(T)$ regimes. Here, we show that, once we have fixed the impurity potential strength in such a way to have an almost pure $\pi$-JJ in the $\alpha=0$ configuration (Fig.\ref{Fig_5_alpha_freccia} (a)), by adding the spin-orbit interaction and modifying its coupling strength, we manage to drive the junction toward $0-\pi$ (Figs.\ref{Fig_5_alpha_freccia} (b-c)) and $0$ (Fig.\ref{Fig_5_alpha_freccia} (d)) regimes. This happens in a reversible manner, in the sense that decreasing the SOC would bring back the system in the initial $\pi$ state.

\section{Pairing functions}\label{correlazioni}

\begin{figure}[h!]
	\includegraphics[scale=0.35]{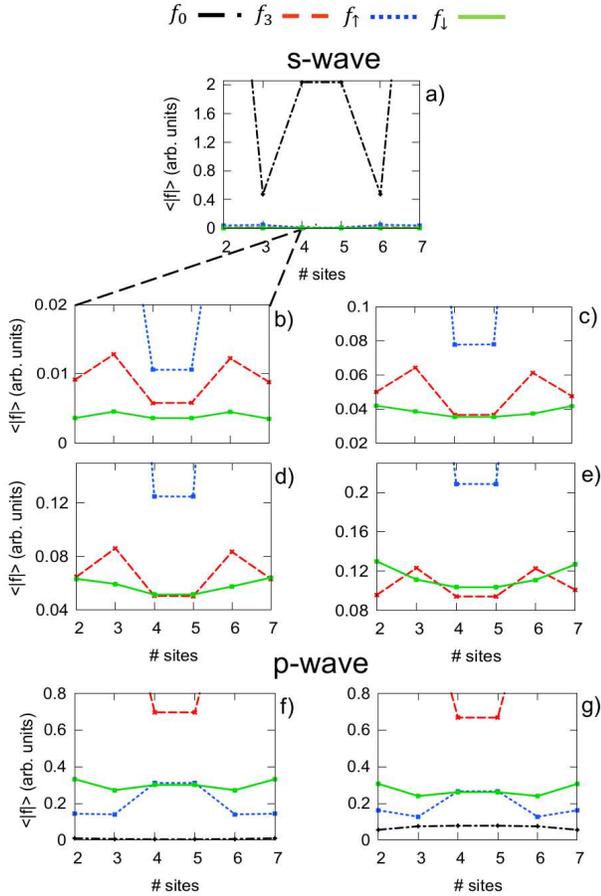}
	\caption[]{Module of spatial profile of s and p-wave pairing components, calculated at $h = 0.45$, $\alpha = 0.04$, $T = 0.025 T_{c}$ and $\phi = 0$, for $V_{imp} = 0.025 $ (a, b, f), $V_{imp} = 0.125 $ (c), $V_{imp} = 0.150$ (d) and $V_{imp} = 0.250 $ (e, g).}
	\label{Fig_6_s_p_wave_B_45_T_0_025}
\end{figure}

A detailed analysis of the pairing mechanisms in these devices is beyond the purpose of this manuscript and will be discussed in detail elsewhere \cite{inpreparation}. However, here, for sake of completeness we give a brief survey of  topic. In SFIS JJs, the exchange field breaks the time-reversal symmetry, giving rise to the zero-spin triplet pairing correlations. However, when we consider systems with impurities and SOC the chance to have more exotic pairing components becomes relevant. In particular, in the presence of SOC, triplet pairings with parallel spins (with $S_{z}= \pm \, 1$ projection) emerge. Indeed, the generation of equal-spin triplet correlations via SOC is provided by the fact that, at interfaces, this latter breaks the spin symmetry, leading to a mixing of spin up and spin down channels in such a way that the total spin $\textbf{S}$ is no longer a good quantum number. As a result, the proximity amplitudes in the ferromagnet will intrinsically be a mixture of singlet and triplet pair correlations. 

Moreover, we notice that SOC breaks also the inversion symmetry at the S/FI interfaces, thus, all four types of pair amplitudes (i.e. s-wave singlet, s-wave triplets, p-wave singlet, p-wave triplets) can be found in the F region \cite{Eschrig2007, Eschrig2008, Lofwander2010}. Since the Pauli's principle requires the pairing correlations to be totally anti-symmetric, the possible types of pairing functions have to fulfill specific symmetry properties with respect to spin, momentum and frequency \cite{Eschrig2007, Eschrig2008, Lofwander2010, Linder_2019, Black_Schaffer_2012, Black_Schaffer_2020, Lothman_2020}. For this reason, the s-wave singlet as well as the p-wave triplets are even functions of the Matsubara frequencies $\omega_{n}$ (even-frequency), while s-wave triplets and p-wave singlet are odd-frequency.

In Figs.\ref{Fig_6_s_p_wave_B_45_T_0_025} we show the amplitude of spatial profile of the calculated s and p-wave correlation functions, as a function of the position inside the FI barrier (expressed in terms of the number $\#$ of sites), corresponding to the system configurations analyzed in the previous section in the presence of SOC and disorder, Figs.\ref{Fig_3_Ic_vimp} (f, g-l). In the following calculations we set $T = 0.025 \, T_{c}$ and $\phi = 0 $. As well as for the systems in Figs.\ref{Fig_3_Ic_vimp} (f, g-l), we choose the following values of the on-site impurity potential: $V_{imp} = 0.025$ (Figs.\ref{Fig_6_s_p_wave_B_45_T_0_025} (a, b, f)), $V_{imp} = 0.125$ (Figs.\ref{Fig_6_s_p_wave_B_45_T_0_025} (c)), $V_{imp} = 0.150$ (Figs.\ref{Fig_6_s_p_wave_B_45_T_0_025} (d)) and $V_{imp} = 0.250$ (Figs.\ref{Fig_6_s_p_wave_B_45_T_0_025} (e, g)). The plots in (b-e) represent a zoomed view of the s-wave spin-triplet components, while in (f, g) we show a zoomed view of the p-wave correlations for the configurations with the lowest and highest value of $V_{imp}$.

In the s-wave symmetry, the majority component is the spin-singlet one $f_{0}$ (for simplicity only shown in Fig.\ref{Fig_6_s_p_wave_B_45_T_0_025} (a)); nevertheless, this is to be expected since we are considering a short-FI barrier directly coupled to conventional s-wave singlet SCs. Furthermore, $f_{0}$ is an even-frequency function and in the Matsubara summation it is reinforced.
We observe that the s-wave spin-triplet pairings, initially generated by SOC at interfaces, survive throughout the FI region and intriguingly appear remarkably enhanced by the effect of increasing the impurity potential $V_{imp}$. 

In the middle of the FI barrier (site $\# \, 4$) we obtain that, passing from $V_{imp} = 0.025$ to $V_{imp} = 0.250$ (Figs.\ref{Fig_6_s_p_wave_B_45_T_0_025} (b, e)), the s-wave singlet $f_{0}$ (even-frequency) remains almost unchanged, while the s-wave spin-triplets (odd-frequency) $f_{3}$, $f_{\uparrow}$ and $f_{\downarrow}$ result increased by a factor $\sim15$, $\sim21$, $\sim33$, respectively.
 
Further, in Figs.\ref{Fig_6_s_p_wave_B_45_T_0_025} (f, g), it is shown that a sizeable p-wave pairing is already induced in the FI layer in the nearly-clean situation ($V_{imp}= 0.025$). In this case, the majority contribution is provided by the zero-spin triplet ($f_{3}$).
However, this latter results enlarged by the influence of non-magnetic disorder, while the triplet correlations (even frequency), contrary to s-wave ones, appear rather stable with respect to the increment of lattice impurities.
Finally, at site $ \# \, 4$ comparing the p-wave pairings with the s-wave ones corresponding to the highest value of impurities strength, we find that the equal-spin components $f_{\uparrow}$ and $f_{\downarrow}$ are almost of the same order of magnitude in both s and p-wave cases.

Our results highlight the importance of SFIS JJs with SOC and tuned impurities as promising platforms hosting unconventional odd-frequency superconductivity and showing sizeable equal-spin triplet pairings, thus verifying the predictions in Refs.\cite{PRLTafuri, ahmad2021coexistence}. 

\section{Conclusion and Discussion} \label{conclusioni}
In this work, we focused our attention on the problem of the tunability of  $0-\pi$ transitions in SFIS JJs. 
The possibility to realize controllable devices that can be switched between different working regimes (namely $0$, $0-\pi$ and $\pi$) would pave the way to the application of JJs with ferromagnetic links in superconducting circuitry \cite{ahmad2021coexistence}. 
Here, we extended the analysis carried out in Ref.\cite{Asano2019} to the case of $0-\pi$ junctions with ferromagnetic insulator barriers. Using a Bogoliubov de Gennes tight-binding model we manage to study the temperature dependent transport properties of these devices. In particular, we studied the influence of spin-mixing and disorder effects on SFIS JJs, focusing on the $I_{c}(T)$ behavior and on the correlation functions, thus deepening the study carried out in Ref.\cite{ahmad2021coexistence}.
We pointed out the role of SOC in driving the switching between $0-\pi$ and $0$ regimes and the capability to induce $0-\pi$ to $\pi$ conversions by adding disorder to the system. 
In particular, the engineering of the impurity concentration (that is strongly linked with the model parameter $V_{imp}$) could lead to the realization of stable $\pi$ junctions, highly desired for superconducting circuits.
 
Moreover, we figure out the opportunity to obtain a fully tunable system,  starting from a $\pi$-JJ and tuning the spin-orbit field by external means. Tuning the SO-interaction in semiconducting quantum wells \cite{SO_tunability,SO_tunability1,Bercioux_2015} can be achieved by gating the structure. In the devices studied here, this procedure is yet unexplored, but, as we have shown, could produce potentially interesting effects and it worths further investigation. 
In this context, the SFIS JJs analyzed here could represent an intriguing and unexplored platform which can be switched among the three different regimes.

 Finally, we complete our analysis by studying the correlation functions in the presence of SOC and impurities. In particular, we observe an enhanced contribution of the odd-frequency pairings, i.e. s-wave triplets and p-wave singlet, due to the increasing of non-magnetic disorder. Therefore, we recognize these tunable SFIS JJs as good candidates to host unconventional superconducting pairing mechanisms and source of sizeable spin-triplet superconductivity, confirming the results of Refs. \cite{ahmad2021coexistence, Asano2019}.

\begin{acknowledgments}
	Financial support and computational resources from MUR, PON “Ricerca e Innovazione 2014-2020”, under Grant No. "PIR01\_00011 - (I.Bi.S.Co.)" are acknowledged. The authors acknowledge Gianluca Passarelli for his invaluable support during the early stage development of the numerical code and A. Tagliacozzo, D. Massarotti and F. Tafuri for fruitful discussions.
\end{acknowledgments}

\subsection*{Appendix A: Tuning of $0$, $0-\pi$ and $\pi$ regimes in SFIS JJs in the clean limit, by varying the exchange field $h$}
\begin{figure}[h!]
	\includegraphics[scale=0.35]{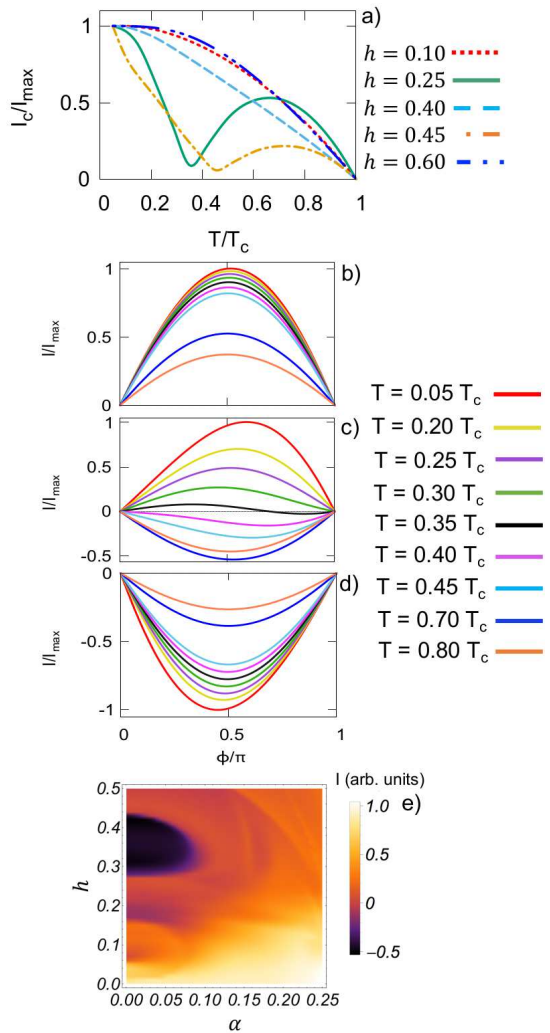}
	\caption[]{(a): $I_c(T)$ for different values of $h$. For these simulations we used $\alpha = 0.04$, $L = 8$ and $W = 32$ sites in the $x$ and $y$ direction. CPR of a $0$ (b), $0-\pi$ (c) and $\pi$ (d) JJ, obtained for $h= 0.10, \, 0.25, \, 0.40 $, respectively. In (e) the Josephson current (at $T=0.1T_{C}$ and $\phi=\pi/2$) as a function of the exchange field $h$ and SOC strength $\alpha$ is shown.}   
	\label{Fig_7_Appendix_A}
\end{figure}

To be thorough, in this section we illustrate how the $I_c(T)$ behavior is affected by varying the exchange field $h$, showing that our model recovers the possibility to switch between the $0$, $0-\pi$ and $\pi$ regimes in SFIS JJs, as reported in the literature \cite{Kontos2002, Blum2002, Robinson2006, Ryazanov2001, Sellier2004, Frolov2004}.

In Fig.\ref{Fig_7_Appendix_A} (a) we show the temperature dependence of the critical current, obtained for increasing values of $h$ in the clean limit. For these simulations we fixed: $\alpha = 0.04$, $\delta_{h}=0$, $V_{imp} = 0$, $L = 8$ and $W = 32$ sites in the $x$ and $y$ direction, respectively. The other parameters are the same as those used in the main text. 

The nonmonotonic dependence of $I_{c}(T)$ is visible. In particular, we can notice that, starting from the AB-type behavior ($h = 0.10$, red curve), $I_{c}$ is strongly modified with increasing the exchange field $h$ in the ferromagnetic layer, causing the system to move towards $0-\pi$ and $\pi$ regimes.

 Then, undergoing some oscillations, the AB trend is established again when $h$ increases ($h = 0.60$, blue curve). Moreover, the cusp-minimum of the $0-\pi$ transition appears to be shifted in temperature as $h$ varies.

We also report in Figs.\ref{Fig_7_Appendix_A} (b-d) the CPR corresponding to the $I_{c}(T)$ curves obtained for $h= 0.10, \, 0.25, \, 0.40 $, respectively. As can be seen, the AB-type curves shown in (a) correspond to pure $0$ (b) and $\pi$ (d) JJs, stable over the whole range of temperatures. 

To better illustrate what we explained in terms of the $I_{c}(T)$ behavior, in Fig.\ref{Fig_7_Appendix_A} (e) we additionally report the density plot of Josephson current (at $T=0.1T_{c}$ and $\phi = \pi/2$) as a function of the exchange field $h$ and SOC strength, $\alpha$, for a wider range of these latter parameters. The black/dark-violet regions indicate negative values of the current, where the JJ is $\pi$. In particular, we can notice a well-defined $\pi$\emph{-island} that in our simulations appear stable in temperature. The white/light-yellow regions, where $I > 0$, indicate conventional $0$-JJs, with a mostly constant occurrence at low values of the exchange field ($h < 0.05 $) for any value of $\alpha$. Intermediate orange-red areas represent JJs which show a good probability to undergo a $0-\pi$ transition, since the critical current is very low. Our results, obtained for SFIS JJs, are in accordance and extend those previously demonstrated in \cite{Asano2019, Asano2001} in the case of SFS JJs in the non-insulating regime.

However, it is worth noting that varying $h$ significantly influences the $I_{c}(T)$ behavior of these systems, producing an even more considerable effect in the short junction regime, namely the situation we are considering in this work. For this reason, manipulating the exchange field of the barrier is effectively a difficult operation to manage experimentally. Conversely, the $0$, $0-\pi$ and $\pi$ switching can be more easily attained acting on the  SOC and non-magnetic impurities.

\subsection*{Appendix B: Recursive Green's function method (RGF)}
The recursive Green's function method (RGF) is a well established technique firstly introduced in the study of electronic transport in mesoscopic systems \cite{Ando_1991, Datta_1995, Lee_1981}. Moreover, in the last twenty years this calculation technique has been widely employed to investigate the transport properties of superconducting Josephson junctions (JJs) \cite{Furusaki1994, Asano2001, Asano2019}. 
Even though the RGF method has been in-depth analyzed and presented in a clear and complete manner in Refs.\cite{Lee_1981, Ando_1991, Furusaki1994, Asano2001, Datta_1995, Asano2019}; in this section, we briefly illustrate the main features of the RGF technique as applied to our case, without the purpose of providing a thorough explanation of the method.

RGF is a sophisticate numerical technique that allows calculating the Green's function (GF) of a central device when it is connected to two leads. This method is extremely useful if the device GF cannot be computed from the direct inversion of its Hamiltonian. 
The RGF can be applied to systems (devices and leads) which are described by lattice models. 

When dealing with a JJ, the two leads are the superconducting electrodes and the central device is the barrier. 

\begin{figure}
	\includegraphics[scale=0.29]{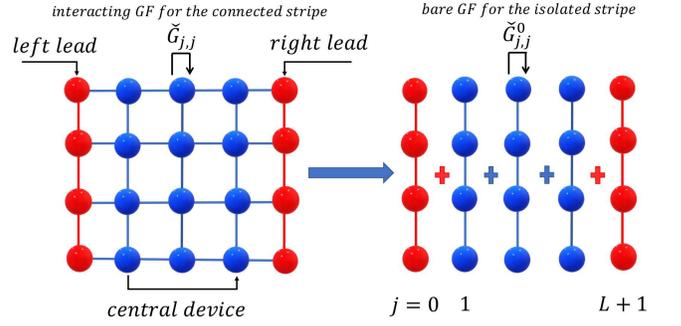}
	\caption[]{Scheme of the system made up of two leads and one
		central device, on the left. On the right, the system divided in stripes. $\check{G}_{j, j}$ is the \textit{interacting} GF of the $j$-th stripe when connected to the others.
		$\check{G}^{0}_{j, j}$ stands for the \textit{bare} GF of the isolated $j$-th stripe.
	}   
	\label{Fig_8_Appendix_B_1}
\end{figure}

Let us specify the discussion to the case analyzed in the main text (see Fig.\ref{Fig_1_model_JJ}):  the system is characterized by a nearest neighbors description, whose tight-binding Hamiltonian is reported in Eqs.(\ref{H_junction}, \ref{H_k_S} - \ref{H_i}).
The RGF method consists in dividing the barrier Fig.\ref{Fig_8_Appendix_B_1}, along the transport direction, in transverse stripes, whose GFs have to be computed by recursively connecting them to the leads. 
In particular, we suppose that the \textit{bare} Matsubara GF of each stripe $j$ can be calculated as follows:
\begin{gather}
\label{bare_GF}
\check{G}^{0}_{j, j}=\left[ i \omega_{n} \check{1} - \check{H}^{0}_{j, j}\right]^{-1} \, ,
\end{gather}
where $\check{H}^{0}_{j, j}$ is the Hamiltonian of the stripe, involving only couplings between lattice sites within the same stripe (i.e. $\hat{H}(\vec{r},\vec{r}')$ with $\vec{r}=j\vec{x}+m\vec{y}$ and $\vec{r}'=j\vec{x}+m^{'}\vec{y}$ in Eq.(\ref{H_junction})).

Then we have to compute the \textit{interacting} GF of the stripe $j$, $\check{G}_{j, j}$ in Eq.(\ref{Nambu_spin_GF_4W}), which corresponds to the GF when it is connected to the two leads and to its nearest neighbors (i.e. $j-1$-th and $j+1$-th stripes).

We observe that $\check{G}_{j, j}$ can be numerically computed by the means of a \textit{Dyson-like} equation \cite{Asano2019, Furusaki1994, Datta_1995}, starting from the surface GFs of the superconductors, which are supposed to be known. These latter are defined as $\check{G}^{L}_{0,0}(\omega_{n})$ and $\check{G}^{R}_{L+1,L+1}(\omega_{n})$ for the left and right S leads (respectively with $j=0$ and $j=L+1$ in Fig.\ref{Fig_1_model_JJ}), and can be evaluated by applying the calculation method used in \cite{Furusaki1994, Ferry2009:book}. 

Here and in the followings we indicate with $\check{G}_{j, j}$ the GF of the $j$-th barrier stripe when it is linked to both the left and right leads, whereas $\check{G}^{L}_{j, j}$ ($\check{G}^{R}_{j, j}$) stands for the GF of the stripe $j$ when it is only connected to the left (right) lead.

For the sake of clarity, we show how to calculate the \textit{interacting} GF for the first stripe of the barrier (with $j=1$, Fig.\ref{Fig_1_model_JJ}) when it is attached to the left S lead (with $j=0$, Fig.\ref{Fig_1_model_JJ}), i.e. $\check{G}^{L}_{1, 1}$ .
Since we are considering only nearest neighbor hoppings, two adjacent stripes can be attached by using the hopping matrix $\check{T}^{\pm}$ in Eq.(\ref{T_matrix}). 

Therefore, we can easily find a \textit{Dyson-like} equation for the first barrier stripe, $\check{G}^{L}_{1, 1}$, starting from its \textit{bare} GF,  $\check{G}^{0}_{1, 1}$: 
\begin{center}
	\begin{align}
	\label{G_11}
	\check{G}^{L}_{1, 1} =  \check{G}^{0}_{1, 1} + \check{G}^{0}_{1,1}\check{T}^{-}_{1,0}\check{G}^{L}_{0, 1}  \, ,
	\end{align}
\end{center}
where we exploit the fact that $\check{T}^{-}_{1,0}$ is the interaction term between the $0$-th (left lead) and the first stripe (barrier).

We observe that this expression for $\check{G}^{L}_{1, 1}$ in Eq.(\ref{G_11}) depends on the knowledge of the GF connecting the stripes $0$ and $1$ (namely $\check{G}^{L}_{0, 1}$), which needs to be evaluated. 
Analogously to $\check{G}^{L}_{1, 1}$, we can derive $\check{G}^{L}_{0, 1}$ as follows :
\begin{center}
	\begin{align}
	\label{G_01}
	\check{G}^{L}_{0, 1} = \check{G}^{L}_{0,0}\check{T}^{+}_{0, 1} \check{G}^{L}_{1, 1}  \, ,
	\end{align}
\end{center}
where we exploit the fact that $\check{G}^{0}_{0, 1}$ vanishes, because the \textit{bare} GFs, $\check{G}^{0}$, cannot connect together different stripes. 
Hence, we obtain two coupled equations for the $\check{G}^{L}_{1, 1}$, Eqs. (\ref{G_11}, \ref{G_01}), that depend only on the \textit{bare} GF of the stripe $1$, the surface GF of the left S lead and the matrices $\check{T}^{\pm}$.
Since the hopping matrices between nearest neighbor stripes (i.e. $\check{T}^{+}_{j,j+1}$ and $\check{T}^{-}_{j+1,j}$) are equal for every value of the index $j$, from now on we simply indicate with $\check{T}^{\pm}$ the stripes interactions. 

\begin{figure}
	\includegraphics[scale=0.30]{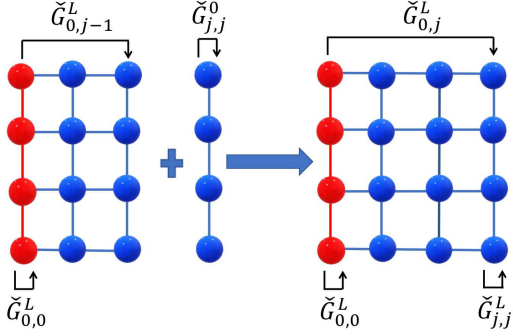}
	\caption[]{Scheme describing how to connect the stripe $j$ to the left lead.
		The first $j-1$ stripes are already connected. In order to add the stripe $j$, $\check{G}^{L}_{j, j}$ and $\check{G}^{L}_{0, j}$ need to be computed.
	}   
	\label{Fig_9_Appendix_B_2}
\end{figure}

With straightforward calculations, we manage to find a more compact expression for $\check{G}^{L}_{1, 1}$:
\begin{gather}
	\check{G}^{L}_{1, 1} =  \left[ i \omega_{n} \check{1} - \check{H}^{0}_{1, 1} - \check{T}^{+} \check{G}^{L}_{0, 0}\check{T}^{-}  \right]^{-1},
\end{gather}
where we explicitly use the Eq.(\ref{bare_GF}). 
This equation can be easily generalized to the $j$-th stripe as follows
\begin{gather}
\label{GL_GR_Asano}
\check{G}^{L}_{j, j} =  \left[ i \omega_{n} \check{1} - \check{H}^{0}_{j, j} - \check{T}^{+} \check{G}^{L}_{j-1, j-1}\check{T}^{-} \right]^{-1},\;0 \leq j \leq L \, ,
\end{gather}
thus, allowing the calculation of the $j$-th stripe GF from the one of the $j-1$-th, Fig.\ref{Fig_9_Appendix_B_2}.

By repeating the procedure, starting from the right (R) side (i.e. starting from the stripe with $j=L+1$ in Fig.\ref{Fig_1_model_JJ}), we can connect the stripe $j$ to the right lead in the following way:
\begin{gather}
	\check{G}^{R}_{j, j} = \left[ i \omega_{n} \check{1} - \check{H}^{0}_{j, j} - \check{T}^{-} \check{G}^{R}_{j+1, j+1}\check{T}^{+}  \right]^{-1}\;L+1 \leq j \leq 1 \, .
\end{gather}

Then, the GF $\check{G}_{j, j}$, in Eq.(\ref{Nambu_spin_GF_4W}), inside the barrier ($ 1 \leq j \leq L $) is computed by using the formula
\begin{gather}
	\label{Gjj_Asano}
	\check{G}_{j, j} = \left[ i \omega_{n} \check{1} - \check{H}^{0}_{j, j} - \check{T}^{-} \check{G}^{R}_{j+1, j+1}\check{T}^{+} \right. \\ \notag
	\left. - \check{T}^{+} \check{G}^{L}_{j-1, j-1}\check{T}^{-}  \right]^{-1}, 
\end{gather}
from which we derive the superconducting pairing correlations with even parity symmetry, in Eq.(\ref{s_wave_correlations}).
At last, we obtain GFs connecting adjacent stripes (namely the stripe $j$ with the ones at $j-1$ and $j+1$) by the relations
\begin{center}
	\begin{align}
	\label{GF_off_diagonal_Asano}
	\notag
	\check{G}_{j+1,j} &=   \check{G}^{R}_{j+1, j+1} \, \check{T}^{+}_{j+1, j} \, \check{G}_{j, j} , \\\notag
	\check{G}_{j,j+1} &=   \check{G}_{j, j} \, \check{T}^{-}_{j, j+1} \, \check{G}^{R}_{j+1, j+1} , \\ 
	\check{G}_{j-1,j} &=   \check{G}^{L}_{j-1, j-1} \, \check{T}^{-}_{j-1, j} \, \check{G}_{j, j} , \\ \notag
	\check{G}_{j,j-1} &=   \check{G}_{j, j} \, \check{T}^{+}_{j, j-1} \, \check{G}^{L}_{j-1, j-1} ,  \\ \notag
	\end{align}
\end{center}
that we use in the calculation of the Josephson current, Eq.(\ref{Josephson_curr}), and odd-parity superconducting correlations, Eq.(\ref{p_wave_correlations}).

\end{document}